# Impact of Geometry on 1D Molecular-Kinetics Simulations of Acoustic-Gravity Wave Propagation into the Exosphere

Jose A. Perez Chavez[1], Orenthal J. Tucker[2], Shane R. Carberry Mogan[3], Robert E. Johnson[4], Christopher Blaszczak-Boxe[1,5]

[1]Howard University Program for Atmospheric Science (HUPAS), Howard University, Washington, D.C. 20059, USA
[2]Department of Atmospheric and Planetary Sciences, Hampton University, Hampton, VA 23668, USA
[3]Royal Institute of Technology, Stockholm, Sweden
[4]Material Science and Engineering, University of Virginia, Charlottesville, VA 22904, USA
[5]Department of Earth, Environment and Equity, Howard University, Washington, D.C. 20059, USA

*Correspondence to*: Jose Perez Chavez (jose.perezchavez@bison.howard.edu)

## Abstract

Direct Simulation Monte Carlo (DSMC) calculations of acoustic gravity wave propagation into the exobase region of a Mars-like atmosphere reveal that radial geometry can reduce wave-driven heating compared to a Cartesian model. We examine two acoustic wave (AW) modes with periods of 11 minutes (AW1) and 5.5 minutes (AW2) propagating from 100 to 320 km altitude using a radial molecular kinetics model. The wave-driven heating was reduced by 40–56% with cycle-averaged temperature gradient $\langle dT/dr \rangle$ decreasing from 9.4 K per scale height $H_0$ to 5.6 K/$H_0$ for AW1 and from 4.4 K/$H_0$ to 1.9 K/$H_0$ for AW2 when accounting for planetary curvature. While the growth in wave density amplitude was attenuated for the 1D radial geometry as well, the heating differences are more pronounced, with both effects driven by geometric spreading accumulating as waves propagate into increasingly rarefied regions. These findings suggest that accounting for curvature effects is crucial when conducting DSMC estimates of acoustic wave contributions to thermospheric heating and atmospheric escape, as Cartesian-based derived counterparts may be overestimated by factors of 1.7–2.3 for these frequencies.

# 1. Introduction

Atmospheric waves, including gravity waves (GWs) are ubiquitous features of all stably stratified planetary atmospheres. They are crucial for understanding the energy and momentum budget of planetary upper atmospheres (England et al., 2017). These waves act as a vertical conveyor belt, transporting energy and momentum between the lower atmospheric layers to higher ones (Medvedev and Yiğit, 2019; Williamson et al., 2019; Yiğit and Medvedev, 2015). GWs are characterized by spatial scales of tens to hundreds of kilometers and periods of minutes to hours. Wave-like perturbations arise from diverse generating mechanisms, including diurnal thermal gradients, interactions with topographic features, and meteorological disturbances such as dust storms (Creasey et al., 2006a; Medvedev et al., 2015, 2011b, 2011a; Wright, 2012; Yiğit, 2023; Yiğit et al., 2021). Recent observational data from the Mars Atmosphere and Volatile Evolution (MAVEN) mission has revealed pronounced density fluctuations propagating into the rarefied regions of the Martian upper atmosphere (England et al., 2017; Terada et al., 2017; Williamson et al., 2019; Yiğit et al., 2015), which have implications for thermospheric heating and cooling effects, exospheric escape, and consequently, the long-term atmospheric evolution (Yiğit, 2021).

The detection history of Martian gravity waves spans multiple decades and mission architectures. Initial evidence emerged from entry accelerometer measurements during the Mars Pathfinder mission (Magalhães et al., 1999), followed by corroborating observations from aerobraking maneuvers of both Mars Odyssey spacecraft (Fritts et al., 2006; Tolson et al., 2007; Withers, 2006) and Mars Global Surveyor (MGS; Ando et al., 2012; Creasey et al., 2006b). Subsequent observations further validated the prevalence of these oscillatory features: wave structures, likely due to thermal tides, were observed in the upper atmosphere (60-130 km) via stellar occultation using the European Space Agency's Mars Express SPICAM (Spectroscopy for the Investigation of the Characteristics of the Atmosphere of Mars; Forget et al., 2009); and vertical wave structures, assumed to be GW in the lower atmosphere (10-30 km), were inferred from MGS radio occultations (Creasey et al., 2006a). More recent investigations using MAVEN's suite of instruments have identified wave-like perturbations in multiple atmospheric constituents, including $CO_2$, O, and Ar (England et al., 2017; Siddle et al., 2019; Wang et al., 2020; Williamson et al., 2019; Yiğit et al., 2015). Observed relative density perturbations of 20-40% show variations with local time, latitude, and altitude (Yiğit et al., 2015). Wave responses vary by species: $CO_2$ and Ar show almost identical density and phase signatures, whereas lighter gases, such as $N_2$ and O, typically exhibit smaller amplitudes and noticeable phase shifts. Notably, some of these perturbations exceed 40% and persist into the exosphere (England et al., 2017; Williamson et al., 2019). Horizontal wavelengths have been characterized as ranging from approximately 100-500 km (Terada et al., 2017). Martian dust storms have the potential to amplify wave activity by a factor of two, doubling hydrogen escape fluxes (Yiğit et al., 2021).

The analytical framework for studying wave propagation in the Martian atmosphere has traditionally relied on linearized hydrodynamic formulations (e.g., England et al., 2017; Parish et al., 2009; Roeten et al., 2022; Walterscheid et al., 2013) or general circulation models incorporating parameterized wave physics (Forbes et al., 2002; Forbes and Hagan, 2000; Forget et al., 1999; Yiğit et al., 2008). While these approaches can provide reasonable approximations for estimating escape rates (Yiğit et al., 2021) and heating rates (Johnson et al., 2021), they encounter a fundamental limitation when applied to the exobase region, where the Knudsen number – the ratio of local mean free path to atmospheric scale heights – approaches and exceeds unity. In this region, the gas transitions to a rarefied state where non-equilibrium effects become predominant and the continuum assumption inherent in fluid dynamics models becomes invalid (Tucker et al., 2016; Volkov et al., 2011).

To accurately describe the tenuous region near an atmosphere's exobase, solutions to the Boltzmann equation or molecular kinetic simulations, such as the Direct Simulation Monte Carlo (DSMC) method (Bird, 1994), have proved essential. DSMC models have been successfully applied to diverse planetary atmospheres, coupling to fluid models of Pluto's lower atmosphere (Erwin et al., 2013; Johnson et al., 2013; Tucker et al., 2012), the Moon's early volcanic atmosphere (Tucker et al., 2021), heating and cooling in multi-component atmospheres induced by escaping $H_2$ (Carberry Mogan et al., 2020, 2021b, 2025; Evans et al., 2025; Tucker et al., 2013), and thermal and non-thermal processes in Europa's atmosphere (Carberry Mogan et al., 2024; Shematovich et al., 2005). Additionally, these models have proven valuable for simulating transient phenomena such as the eclipse-induced collapse and subsequent reformation of Io's and Ganymede's sublimated atmospheric components (Milby et al., 2024; Moore et al., 2009), and plumes on Enceladus and Europa (Berg et al., 2016; Dayton-Oxland et al., 2023; Mahieux et al., 2019; Tseng et al., 2025, 2022; Tucker et al., 2015).

The DSMC method has been used to calculate the wave-induced density and temperature perturbations, as well as their relative phases (Leclercq et al., 2020; hereafter L20), that may invalidate existing temperature profiles derived from MAVEN density measurements under the assumption of hydrostatic equilibrium. Similarly, the method directly captures heating processes of sustained wave activity (Tian et al., 2023; hereafter T23). This calculation is possible in conventional linear fluid models only via second-order calculations to account for heating effects (Johnson et al., 2021; Schubert et al., 2005; Walterscheid et al., 2013). Whereas prior DSMC studies used 1D Cartesian (i.e., "slab") geometries, the model used in this study prescribes a spherically symmetric domain and adequately describes the inherent curvature of planetary atmospheres. The work presented here examines the impact of geometry on AWs propagation in a Martian-like upper atmosphere using the DSMC method by comparing wave propagation characteristics and atmospheric response from prior 1D Cartesian to our more realistic 1D radial model.

The remainder of this paper is organized as follows: Section 2 details the DSMC implementation using a spherically symmetric domain, including boundary conditions, numerical parameters, and analysis techniques. Section 3 presents our findings on wave behavior, amplitude growth, and temperature gradients, comparing 1D radial and Cartesian geometries for two acoustic wave modes in T23. Section 4 concludes with a discussion of the implications for atmospheric modeling and interpretation of spacecraft.

## 2. Methods

### 2.1 Description of the model

The DSMC method does not rely on continuum fluid assumptions and instead solves the Boltzmann equation statistically, allowing it to capture the physics across the transition from collisional to collisionless regimes. A 1D radial DSMC model with a spherically symmetric geometry is employed here, following prior applications to the tenuous, water-product atmospheres on the icy Galilean satellites (Carberry Mogan et al., 2025, 2024, 2022, 2021b, 2021a, 2020; Waite et al., 2024) as well as multi-component planetary upper atmospheres (Evans et al., 2025). Here we considered a single-species O atmosphere with absent background wind, which allows us to compare to T23. That is, we use identical physical characteristics to quantify how spherical treatment modifies wave propagation characteristics before extending to more complex multi-species atmospheres in future work.

The method tracks computational particles representing O atoms, through 3D space subject to gravity and intermolecular collisions. Atmospheric properties including density, $n(i)$, and temperature, $T(i)$, are calculated directly from the particle statistics within each radial cell,

$$n(i) = \frac{N(i)w}{V_i}, \qquad T(i) = \frac{m}{3k_B}\left(\frac{\sum_{p=1}^{N(i)} v_p^2 w}{n(i)V_i} - \langle v_p(i)\rangle^2\right), \tag{1}$$

where $N(i)$, $w$, $V_i$, and $\langle v_p(i)\rangle$ are the particle count, statistical weight, cell volume, and mean velocity in cell $i$, respectively.

The domain extends from 100–700 km in altitude divided into 80 radially varying cells with finest resolution (1.5 km) near the lower boundary. The geometry is bound by concentric inner ($r_0$) and outer ($r_1$) spheres above the Martian surface ($R_M = 3396$ km), where lower and upper boundary conditions are applied (Table 1). Collisions between O atoms are modeled using a hard-sphere cross-section of $\sigma_{xsec} = 4.5 \times 10^{16}$ cm², consistent with L20 and T23.

| ATMOSPHERIC PARAMETERS | Parameter [units] | Symbol | Value |
|---|---|---|---|
| | Number density at $r_0$ [m⁻³] | $n_0$ | $10^{16}$ |
| | Temperature at base [K] | $T_0$ | 270 |
| | Scale height [km] | $H_0$ | 40 |
| | Mean free path at $r_0$ [km] | $\lambda_0$ | 1.50 |
| | Acoustic frequency at $r_0$ [rad/s] | $\omega_a$ | $6.05 \times 10^{-3}$ |
| SIMULATION PARAMETERS | Time step [s] | $\Delta t$ | 0.098 |
| | Number of cells | - | 80 |
| | Total simulation particles | - | $3 \times 10^6$ |
| | Collisional cross-section [cm²] | $\sigma_{xsec}$ | $4.5 \times 10^{-16}$ |
| | Lower radial boundary [km] | $r_0$ | 100 |
| | Upper radial boundary [km] | $r_1$ | 700 |
| | Initialization period [s] | - | 6000 |

*Table 1. Atmospheric parameters as in T23 and simulation parameters representing a simplified Mars-like atmosphere with atomic oxygen as the primary constituent.*

At the lower boundary, number density $n_0$ and temperature $T_0$ are fixed. The upper boundary at 700 km altitude implements a free-escape condition, permanently removing particles that cross this threshold. The validation of our boundary choice is discussed in Carberry Mogan et al., 2021a. Simulations employ approximately three million particles and run for 6000s to achieve steady state before introducing wave perturbations.

As in L20 and T23, perturbations are generated via sinusoidal modulation of the upward particle flux $F(t)$ at the lower boundary as a function of time $t$:

$$F(t) = F_0[1 + A\sin(\omega t)], \qquad (2)$$

where $F_0 = \frac{1}{4}n_0\langle v \rangle$ is the equilibrium upward flux, $\langle v \rangle = \sqrt{\frac{8k_B T_0}{\pi m}}$ is the average thermal speed obtained from the Maxwell Boltzmann distribution. The amplitude of the perturbation $A$ is relative to $F_0$, and $\omega$ is the wave frequency. Two modes are examined: AW1 ($\omega_1 = 9.5 \times 10^{-3}$ rad/s) and AW2 ($\omega_2 = 19 \times 10^{-3}$ rad/s) as outlined in Table 2. The two modes are chosen directly from T23 for comparison and driven by Eq. (2). Both wave frequencies are above the acoustic cut-off frequency ($\omega_a = 6.05 \times 10^{-3}$) and well below the Nyquist frequency of $\omega_{Nyquist} = 92.3 \times 10^{-3}$ rad/s for the 34 s sampling interval implemented here providing adequate temporal resolution. This approach allows direct isolation of geometric effects by comparing radial and Cartesian acoustic wave propagation (T23). Internal gravity waves, which require horizontal as well as vertical structure, cannot be simulated in a purely 1D radial framework.

**WAVE PARAMETERS**

| Parameter | Symbol | Value |
|---|---|---|
| Frequency, AW1 [rad/s] | $\omega_1$ | $9.5\times 10^{-3}$ |
| Vertical acoustic wavelength AW1 [km] | $\lambda_1$ | 414 |
| Period AW1 [min] | $\tau_1$ | 11 |
| Frequency AW2 [rad/s] | $\omega_2$ | $19.0\times 10^{-3}$ |
| Vertical acoustic wavelength AW2 [km] | $\lambda_2$ | 169 |
| Period AW2 [min] | $\tau_2$ | 5.5 |
| Flux amplitude at $r_0$ | $A$ | 0.25 |

*Table 2. Wave parameters used here and in T23 for both acoustic wave modes. The acoustic wavelengths $\lambda_1$ and $\lambda_2$ are estimated as in T23 via the full dispersion relation under the assumption that the horizontal wavelength is much larger than the vertical wavelength (Hines, 1960).*

The mean free path $\lambda_{mfp}$ increases with the atmosphere becoming more rarefied at higher altitudes, reaching values of ~37km at the nominal exobase $r = 230$km, and ~176km at an altitude of $r = 300$km. For most of the domain below 300km, the acoustic wavelengths are larger than the $\lambda_{mfp}$ (see Table 2). In the rarefied upper regions, wave-like perturbations are manifested through collective modulation of particle velocities and densities rather than through classical fluid coupling. The flux modulation at the lower boundary (via Eq. 2) launches alternating compressional and rarefactional fluxes of particles that carry correlated momentum and energy upward. These particle ensembles, even when partially ballistic, produce oscillatory variations in density, temperature, and velocity fields, as will be seen in Section 3.

## 2.2 Analysis method

Wave properties are computed using additional analysis steps, following the methodology of T23:

1. Perturbation amplitudes: The density and temperature data, calculated as in Eq. (1), are normalized against the isothermal steady-state values to get amplitudes relative to the unperturbed atmosphere, $\frac{n(r)-n_0}{n_0}$ and $\frac{T(r)-T_0}{T_0}$.
2. Amplitude growth: The wave amplitude peaks are tracked in time and altitude to examine the effect of the perturbation with altitude, time to steady state and dissipation after forcing is ceased.

3. Heating effects: The average temperature increase against the background atmosphere ($\langle T \rangle - 270$ K) is computed over each wave cycle to quantify atmospheric heating. The wave-induced temperature gradient is computed as a function of altitude.

# 3. Results

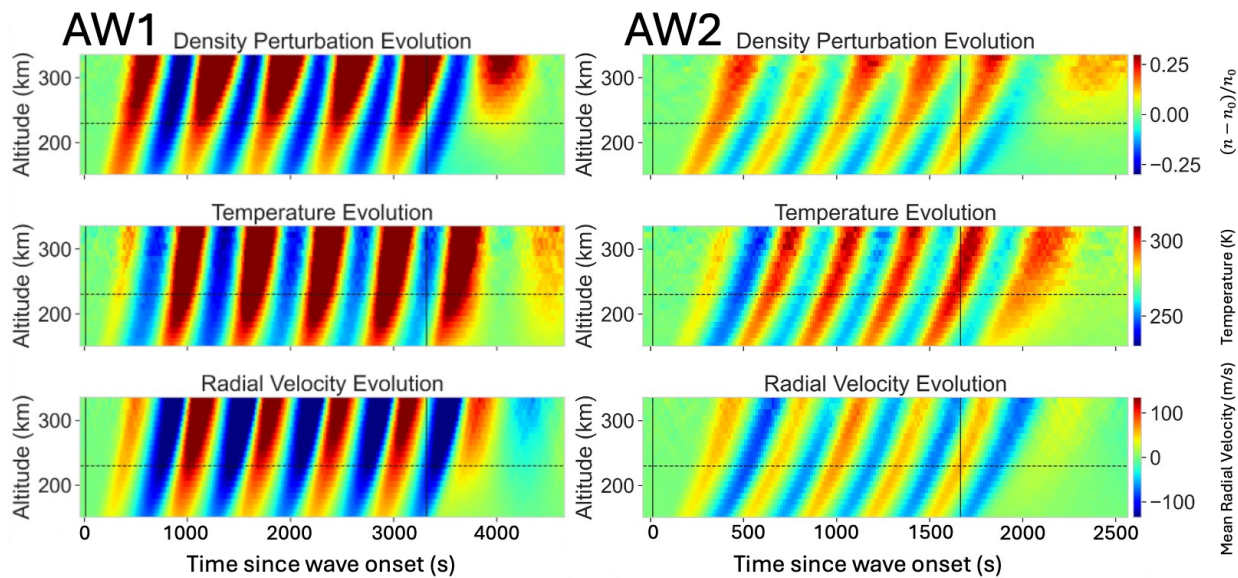

*Fig. 1. Overview of radial and temporal* evolution *of AW modes following wave onset after steady state achieved* for the background isothermal atmosphere *at 6000s. From top to bottom: density normalized amplitudes, temperature, and radial velocity. Left panels: Lower frequency mode of 11-minute period, AW1. Right panels: Higher frequency mode of 5.5-minute period, AW2. The black horizontal dashed line shows the nominal exobase altitude at 230 km. The color bars for each gas apply to both left and right panels. Vertical solid lines delineate the start and end of the wave forcing at the lower boundary (100km).*

The radial and temporal evolution of the two acoustic wave modes, AW1 and AW2, are shown in Fig. 1. The panels show the variation in the atmosphere properties, as computed in Eq. (1), produced by the wave modes following the onset at 6000 seconds. At this time, the particle flux at the lower boundary is oscillated sinusoidally with a dimensionless amplitude of 0.25 via Eq. (2), generating alternating peaks and troughs in density, temperature, and radial velocity indicated by the distinct color bands. To compare to the results of T23, five complete cycles of wave forcing are imposed over approximately 3300 s for AW1 and 1650 s for AW2. The perturbation travels upward through the atmosphere, with clear phase progression visible in all three fields as the wave structure ascends from the lower boundary to beyond the nominal exobase at 230 km. Wave forcing is terminated at the conclusion of the fifth cycle, after which remnants of the perturbation persist throughout the domain, particularly evident in the rarefied exobase region where dissipation occurs more slowly due to reduced collision frequencies. Both wave modes demonstrate increasing amplitude growth in the propagating perturbations in all three fields, typical behavior of AWs. The density amplitude perturbations, normalized against the steady state unperturbed profile $n_0(r)$, are shown on the top panels.

## 3.1. Temporal Evolution of Wave Behavior

Significant differences between the two wave modes in temperature and radial velocity are shown in Fig. 2. Temperature perturbations due to AW1 exceed 300 K at altitudes of ~180 km, whereas AW2 obtains similar temperatures at higher altitudes of ~280–300 km. The contrasting temperature amplitudes are consistent with the higher vertical speeds for AW1 compared to AW2. When wave activity is terminated, activity briefly persists throughout the domain before decaying, particularly in the rarefied exobase region of AW1. These broader differences in the behavior of AW1 and AW2 are consistent with T23.

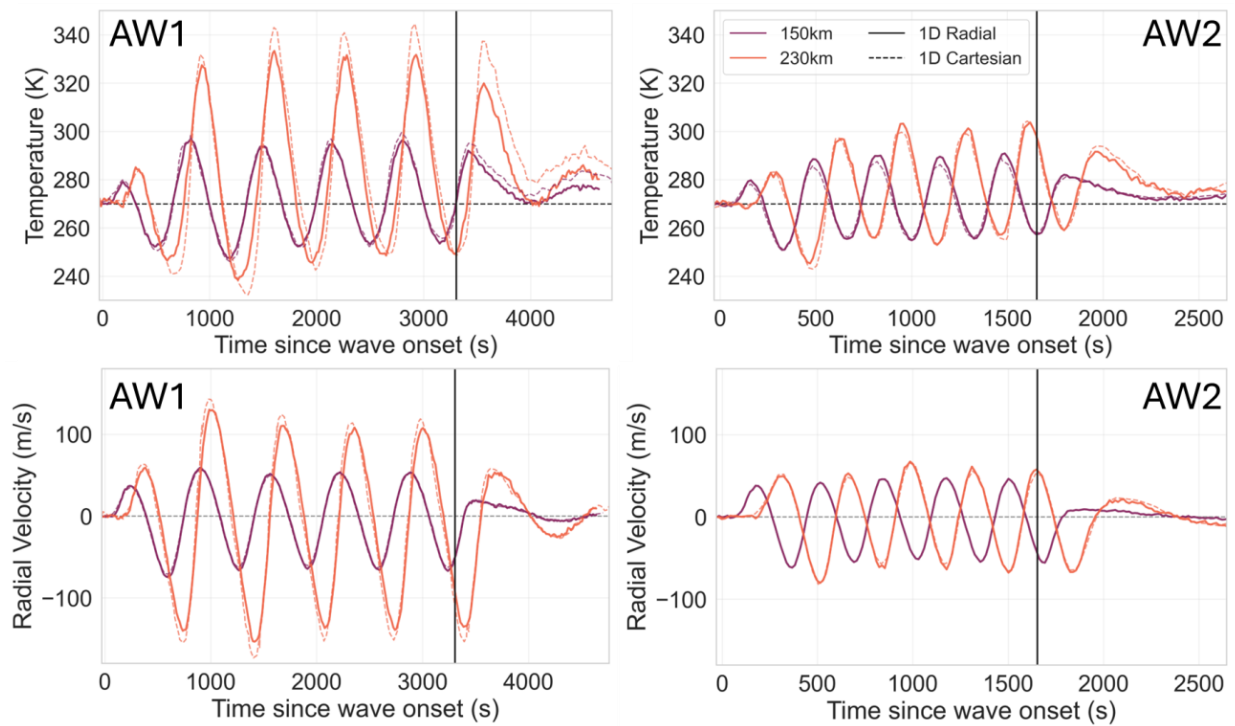

*Fig. 2. Temperature (K) and mean radial velocity (m/s) evolution over time following wave onset and decay for AW1 (left panels) and AW2 (right panels) modes at two altitudes, 150km and 230km. Solid vertical line indicates termination of perturbations. Solid color lines represent data*

*from the 1D radial simulation in this study, while dashed lines correspond the 1D Cartesian results from T23. The horizontal black dashed line in the top panels marks the 270 K isothermal background profile.*

Fig. 2 illustrates the temporal evolution of temperature and mean radial velocity at selected altitudes (150 km, 230 km) for the 2 AW modes. AW1 (left panels, solid lines) exhibits pronounced temperature oscillations with peak-to-trough swings of approximately 45 K at 150 km altitude and nearly 100 K at 230 km. The oscillation amplitude roughly doubles over an 80 km span, a substantial amplification resulting from energy transport and wave growth in the upper atmosphere. The radial velocity perturbations follow a similar pattern, with oscillations of ±60 m/s at 150 km while at 230 km the oscillations display larger swings ranging from −170 m/s to +120 m/s.

In contrast, AW2 (right panels, solid lines in Figure 2) demonstrates different behavior. Temperature oscillations span approximately 40 K at 150 km and 60 K at 230 km. The radial velocity perturbations in AW2 remain more uniform across altitudes, with both 150 km and 230 km exhibiting similar magnitudes of ±50-60 m/s. The vertical amplification of temperature oscillations due to 1D radial geometry are less pronounced with the higher frequency mode AW2 than AW1, as the former increases by 50% while AW1 doubles in temperature amplitude for the same altitude span.

Comparison with T23 results (dashed lines) reveals good agreement for AW2 across both altitudes, with the 1D radial geometry reproducing the 1D Cartesian results closely. For AW1, the lower altitude (150 km) shows reasonable alignment between geometries, but at 230 km, the 1D Cartesian approach overestimates heating effects. Temperature peaks in the simulation presented here are approximately 25 K lower than the 1D Cartesian predictions at this altitude. As the atmosphere responds to wave perturbations, the cycle-averaged temperature gradually shifts from oscillating about the background 270 K value to oscillating about a time-dependent baseline that increases with both time and altitude, indicative of wave-induced heating as discussed further below.

## 3.2. Wave Amplitude Growth

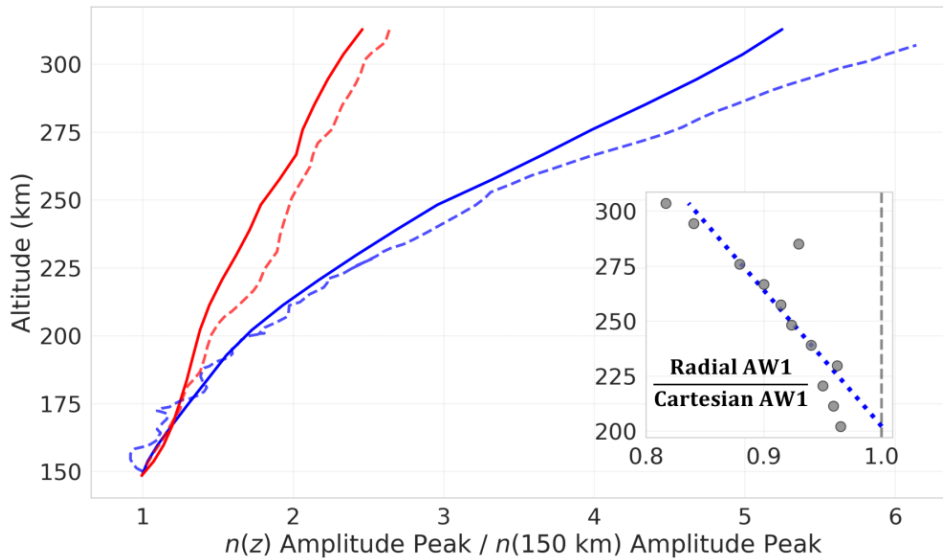

*Fig. 3. Growth in the density amplitude peak versus altitude in the fifth cycle, normalized by the respective amplitude at 150 km. Comparison of the 1D radial geometry results in this study (solid lines) with the 1D Cartesian geometry (dashed lines) for AW1 (blue) and AW2 (red). Inset panel: Ratio of AW1 amplitude growth of the 1D radial to Cartesian geometry showing progressive deviation from unity with altitude. The growth in amplitude peak of AW1 for the 1D radial geometry divided by the growth of AW1 for the 1D Cartesian model along interpolated points and shown as gray points in the inset plot. The blue dotted line is a best-fit line.*

Our analysis examines growth in the density amplitude peak versus altitude during the fifth wave cycle, with amplitudes normalized by their respective values at 150 km (Figure 3). Whereas the differences are small over a scale height ~40km, the lower-frequency AW1 mode exhibits a systematic deviation from Cartesian result that increases with altitude. At the altitude of 300 km, the 1D radial geometry shows a ~20% reduction in amplitude peak growth for AW1 compared to the Cartesian AW1 results, with the amplitude increasing by a factor of ~4.8 from the 150 km reference level versus the Cartesian value of ~5.8. In contrast, the higher-frequency mode AW2 exhibits a roughly constant altitude-independent offset, showing approximately 10% lower values than the Cartesian model with variations of about 5% across different altitudes.

The inset panel of Figure 3 quantifies this geometric influence by showing the ratio of AW1 amplitude peak growth between the 1D radial and Cartesian geometries, clearly illustrating the progressive departure from unity with increasing altitude. This altitude-dependent reduction for AW1 differs with AW2's uniform offset, suggesting that the two modes respond differently to 1D radial geometry. While this progressive divergence with altitude is consistent with a geometric spreading effect, where wave energy is distributed over an expanding cross-sectional area in radial coordinates, the contrasting frequency responses indicate that additional factors beyond pure geometry are contributing. We carried out extended simulations that revealed that AW2 requires more than 5 cycles to converge to a steady propagation differing from AW1, and the fifth-cycle results for AW2 are not in steady state. These simulations are further discussed in Appendix A1.

## 3.3. Temperature Deviations and Heating Effects

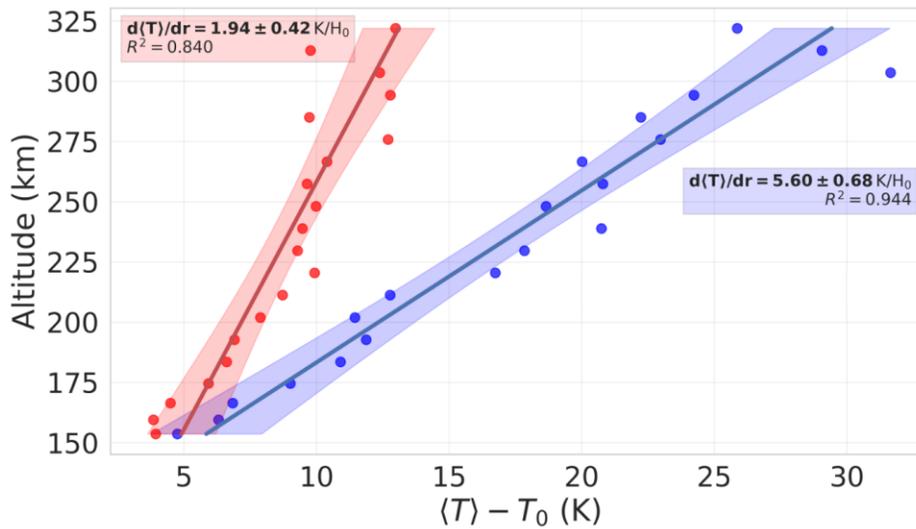

*Fig. 4. Wave-induced cycle-averaged temperature ⟨T⟩ in the fifth cycle minus the isothermal profile with $T_0 = 270$ K (x-axis) as a function of altitude (y-axis). The blue and red colors correspond to AW1 and AW2 modes, respectively, and the solid lines represent the best-fit lines. The slope of each line is the average wave-induced temperature gradient $d⟨T⟩/dr$: 5.60 K/$H_0$ for AW1 and 1.94 K/$H_0$ for AW2. The shaded areas represent the 95% (2σ) confidence band for the best fit.*

Characterizing the oscillation about $T_0$ during wave forcing, the cycle-averaged temperature ⟨T⟩ is computed as in T23. At each altitude level of the domain, the peak-to-peak average temperature during the fifth cycle ⟨T⟩ is computed and subtracted by the isothermal profile of $T_0 = 270$ K, which corresponds to the x-axis Figure 4. The wave produces an overall temperature increase, with higher temperatures deviating away from the isothermal profile with increasing altitude. In Figure 4, the cycle-averaged temperature is seen to grow linearly over the altitude span of 150-320km from ~5 K to ~30 K for AW1, and from ~3 K to 12 K for AW2. This is consistent with the Cartesian results as described in T23, but here we obtain lower cycle-averaged temperature gradients

The 11-minute period AW1 produces a wave-induced gradient $d\langle T\rangle/dr$ of 5.60±0.68 K per $H_0$ scale height, representing a ~40% decrease compared to T23's reported 9.4 K/$H_0$ for the identical wave mode. The 5.5-minute period AW2 shows a lower gradient of 1.94±0.42 K/$H_0$, corresponding to a ~56% reduction relative to T23's 4.4 K/$H_0$ for this higher frequency mode. The average temperature fit line of AW1 has a coefficient of determination $R^2 = 0.94$, a stronger linearity than AW2 with $R^2 = 0.84$, where the smaller signal amplitude in AW2 produces higher scatter. Of course, with much higher frequencies (shorter acoustic wavelengths) the heating effect becomes negligible.

# 4. Discussion & Conclusion

## 4.1 Physical Mechanisms

These DSMC simulations demonstrate that a spherically symmetric geometry reduces both AW-driven heating and amplitude growth in a Mars-like exobase region as compared to the Cartesian models. The cycle-averaged temperature gradient $d\langle T\rangle/dr$ extracted after 5 wave cycles shows reduced values by factors of 1.7-2.3 across both modes. The reduced heating effect arises from geometric spreading inherent to radial coordinates. As waves propagate upward, their energy flux is distributed over an expanding volume following an inverse-square dependence on radius. While the conversion of ordered wave motion to random thermal motion operates in both Cartesian and spherically symmetric geometries, the latter reduces the average wave energy density through geometric spreading.

The cross-sectional area expands by ~10% between 150 and 320 km altitude on Mars. This expansion produces a 19% decrease in amplitude peak growth at an altitude of 300km, with about 10% decrease for AW2. However, our extended simulations (Appendix A1) demonstrate that AW2 requires more cycles, not surprisingly, than AW1 to reach steady propagation unlike what was assumed in T23. Of course, higher-frequency waves are known to dissipate their energy more rapidly during their ascent (Fedorenko et al., 2021; Imamura et al., 2016; Rhode et al., 2024) reducing the effect of curvature.

This attenuation of heating and amplitude growth in the exobase regime for both AWs in the radial geometry underscores the importance of using appropriate geometric frameworks when modeling wave-driven processes in planetary upper atmospheres.

## 4.2 Implications for Mars Atmospheric Science

The findings here have critical implications for understanding energy transport in Mars' exosphere:

**1. Spacecraft Data Interpretation:** The results presented here align with L20's findings of temperature-density phase disparities that invalidate hydrostatic equilibrium assumptions, and T23's demonstration of wave-induced heating effects absent in linear fluid models. The persistence in wave-like activity after the driving perturbation ceases in the exosphere region, as seen in Fig. 1 and 2, underscores how the interpreting spacecraft density data using models that assume steady wave activity can be problematic. Notably, the hotter temperatures linger longer for AW1 than AW2 after oscillations at the lower boundary cease, illustrating the complex interplay between wave frequency, heating mechanisms, and the interpretation of observational data.

**2. Atmospheric Escape Modeling:** Previous estimates of wave-enhanced escape rates require DSMC treatment and accurate geometric representation. For thermal H escape, gravity waves during dust storms, as in Mars, can enhance escape fluxes by factors of 2–5 (Yiğit et al., 2021). However, the likely significant frequency dependent reduction in acoustic wave-driven heating demonstrated here implies correspondingly lower exospheric temperatures in spherical versus Cartesian frameworks. This can directly affect thermal escape calculations of H. For non-thermal escape processes, hot O escape may be primarily enhanced not by altering the escape probability but by increasing the hot O production rate (Gu et al., 2024). Since production rates depend on local density distributions, the altered wave amplitude growth and density perturbations in spherical geometry would affect hot O production estimates.

**3. Energy Budget Calculations:** Both linearized hydrodynamic formulations and global circulation models incorporating wave-driven heating at altitudes nearing the exosphere region based on Cartesian approximations can be problematic. The reduced heating efficiency obtained in this study emphasizes the importance of considering proper geometry of the upper atmosphere when estimating the energy balance in the upper thermosphere.

## 4.3 Scope and Future Directions

This study examines steadily driven acoustic waves in a simplified single-species atmosphere to isolate geometric effects. Although Martian atmospheric waves are highly variable, these controlled conditions allow direct comparison between the different geometries. The 1D approach used here and in earlier DSMC studies is roughly applicable when horizontal wavelengths significantly exceed vertical wavelengths—a condition often met for gravity waves in planetary atmospheres.

The 1D geometry employed here is applied to acoustic wave propagation. Simulating internal gravity waves requires horizontal structure that necessitates a 2D simulation, which are in progress. These simulations will also include $CO_2$ and lighter species to examine species-dependent heating effects and enable quantitative comparison with MAVEN spacecraft measurements. The spherical framework for wave propagation established here provides the foundation for these extensions and is directly applicable to smaller bodies like Pluto and Titan, where curvature effects on wave behavior are expected to be more pronounced in their extended atmospheres.

# Acknowledgments

J.P.C. efforts were supported primarily by the NASA FINESST grant 80NSSC24K1722. Additional support for J.P.C. was in part by NASA MUREP program through the NASA Office of STEM Engagement and the National GEM Consortium via summer fellowships as a 2023-2024 Fellow. The Graduate School at Howard University provided school funds via a GEM Graduate Assistantship during 2023-2024. In addition, the project continued to be supported via Ernest E. Just-Percy L. Julian Graduate Research Assistantship at Howard University in 2024-2025. O.J.T.'s effort was supported through the Exosphere Ionosphere Magnetosphere Modeling GSFC ISFM.  This work used Bridges2 at Pittsburg Supercomputing Center through allocation PHY240179 from the Advanced Cyberinfrastructure Coordination Ecosystem: Services & Support (ACCESS) program, which is supported by U.S. National Science Foundation grants #2138259, #2138286, #2138307, #2137603, and #2138296 (Boerner et al., 2023). We thank L. Tian for providing Cartesian geometry simulation data from Tian et al. (2023) for comparison purposes.

# References

Ando, H., Imamura, T., Tsuda, T., 2012. Vertical wavenumber spectra of gravity waves in the Martian atmosphere obtained from Mars Global Surveyor radio occultation data. J. Atmos. Sci. 69, 2906–2912.

Berg, J.J., Goldstein, D.B., Varghese, P.L., Trafton, L.M., 2016. DSMC simulation of Europa water vapor plumes. Icarus 277, 370–380.

Bird, G.A., 1994. Molecular Gas Dynamics And The Direct Simulation Of Gas Flows.

Boerner, T.J., Deems, S., Furlani, T.R., Knuth, S.L., Towns, J., 2023. ACCESS: Advancing innovation, in: Practice and Experience in Advanced Research Computing. Presented at the PEARC '23: Practice and Experience in Advanced Research Computing, ACM, New York, NY, USA. https://doi.org/10.1145/3569951.3597559

Carberry Mogan, S.R., Moore, L.E., Liuzzo, L., Poppe, A.R., 2025. A direct simulation Monte Carlo–driven photochemical model of Callisto's ionosphere. Planet. Sci. J. 6, 106.

Carberry Mogan, S.R., Poppe, A.R., Liuzzo, L., 2024. The influence of non-thermal collisions in Europa's atmosphere. Geophys. Res. Lett. 51. https://doi.org/10.1029/2024gl109534

Carberry Mogan, S.R., Tucker, O.J., Johnson, R.E., 2021a. The influence of upper boundary conditions on molecular kinetic atmospheric escape simulations. Planet. Space Sci. 205, 105302.

Carberry Mogan, S.R., Tucker, O.J., Johnson, R.E., Roth, L., Alday, J., Vorburger, A., Wurz, P., Galli, A., Smith, H.T., Marchand, B., Oza, A.V., 2022. Callisto's atmosphere: First evidence for $H_2$ and constraints on $H_2O$. J. Geophys. Res. Planets 127. https://doi.org/10.1029/2022je007294

Carberry Mogan, S.R., Tucker, O.J., Johnson, R.E., Sreenivasan, K.R., Kumar, S., 2020. The influence of collisions and thermal escape in Callisto's atmosphere. Icarus 352, 113932.

Carberry Mogan, S.R., Tucker, O.J., Johnson, R.E., Vorburger, A., Galli, A., Marchand, B., Tafuni, A., Kumar, S., Sahin, I., Sreenivasan, K.R., 2021b. A tenuous, collisional atmosphere on Callisto. Icarus 368, 114597.

Creasey, J.E., Forbes, J.M., Hinson, D.P., 2006a. Global and seasonal distribution of gravity wave activity in Mars' lower atmosphere derived from MGS radio occultation data. Geophys. Res. Lett. 33. https://doi.org/10.1029/2005gl024037

Creasey, J.E., Forbes, J.M., Keating, G.M., 2006b. Density variability at scales typical of gravity waves observed in Mars' thermosphere by the MGS accelerometer. Geophys. Res. Lett. 33, L22814.

Dayton-Oxland, R., Huybrighs, H.L.F., Winterhalder, T.O., Mahieux, A., Goldstein, D., 2023. In-situ detection of Europa's water plumes is harder than previously thought. Icarus 395, 115488.

England, S.L., Liu, G., Yiğit, E., Mahaffy, P.R., Elrod, M., Benna, M., Nakagawa, H., Terada, N., Jakosky, B., 2017. MAVEN NGIMS observations of atmospheric gravity waves in the Martian thermosphere. J. Geophys. Res. Space Phys. 122, 2310–2335.

Perez Chavez et al., *Icarus Accepted* (Dec. 2025)Erwin, J., Tucker, O.J., Johnson, R.E., 2013. Hybrid fluid/kinetic modeling of Pluto's escaping atmosphere. Icarus 226, 375–384.

Evans, J.C., Carberry Mogan, S.R., Johnson, R.E., Tucker, O.J., 2025. Thermally driven atmospheric escape: Transition from diffusion-limited to drag-off escape. Planet. Sci. J. 6, 39.

Fedorenko, A.K., Kryuchkov, E.I., Cheremnykh, O.K., Selivanov, Y.A., 2021. Dissipation of acoustic–gravity waves in the Earth's thermosphere. J. Atmos. Sol. Terr. Phys. 212, 105488.

Forbes, J.M., Bridger, A.F.C., Bougher, S.W., Hagan, M.E., Hollingsworth, J.L., Keating, G.M., Murphy, J., 2002. Nonmigrating tides in the thermosphere of Mars. Journal of Geophysical Research: Planets 107, 23-1-23–12.

Forbes, J.M., Hagan, M.E., 2000. Diurnal Kelvin wave in the atmosphere of Mars: Towards an understanding of 'stationary' density structures observed by the MGS accelerometer. Geophys. Res. Lett. 27, 3563–3566.

Forget, F., Hourdin, F., Fournier, R., Hourdin, C., Talagrand, O., Collins, M., Lewis, S.R., Read, P.L., Huot, J.-P., 1999. Improved general circulation models of the Martian atmosphere from the surface to above 80 km. J. Geophys. Res. 104, 24155–24175.

Forget, F., Montmessin, F., Bertaux, J.-L., González-Galindo, F., Lebonnois, S., Quémerais, E., Reberac, A., Dimarellis, E., López-Valverde, M.A., 2009. Density and temperatures of the upper Martian atmosphere measured by stellar occultations with Mars Express SPICAM. J. Geophys. Res. 114. https://doi.org/10.1029/2008je003086

Fritts, D.C., Wang, L., Tolson, R.H., 2006. Mean and gravity wave structures and variability in the Mars upper atmosphere inferred from Mars Global Surveyor and Mars Odyssey aerobraking densities. J. Geophys. Res. 111, A12304.

Gu, H., Cui, J., Wang, X., Huang, X., Zhao, J.-J., Wu, Z.-P., Li, L., 2024. Non-thermal oxygen escape on Mars in the presence of gravity waves. J. Geophys. Res. Space Phys. 129, e2023JA032154.

Hines, C.O., 1960. Internal atmospheric gravity waves at ionospheric heights. Can. J. Phys. 38, 1441.

Imamura, T., Watanabe, A., Maejima, Y., 2016. Convective generation and vertical propagation of fast gravity waves on Mars: One- and two-dimensional modeling. Icarus 267, 51–63.16

Johnson, R.E., Volkov, A.N., Erwin, J.T., 2013. MOLECULAR-KINETIC SIMULATIONS OF ESCAPE FROM THE EX-PLANET AND EXOPLANETS: CRITERION FOR TRANSONIC FLOW. ApJL 768, L4.

Johnson, R.E., Woodson, A.K., Tian, L., Tucker, O.J., Williamson, H.N., 2021. Temperature extraction from spacecraft density profiles in the presence of wave activity. Icarus 357, 114257.

Leclercq, L., Williamson, H.N., Johnson, R.E., Tucker, O.J., Tian, L., Snowden, D., 2020. Molecular kinetic simulations of transient perturbations in a planet's upper atmosphere. Icarus 335, 113394.

Magalhães, J.A., Schofield, J.T., Seiff, A., 1999. Results of the Mars Pathfinder atmospheric structure investigation. J. Geophys. Res. 104, 8943–8955.

Mahieux, A., Goldstein, D.B., Varghese, P.L., Trafton, L.M., 2019. Parametric study of water vapor and water ice particle plumes based on DSMC calculations: Application to the Enceladus geysers. Icarus 319, 729–744.

Medvedev, A.S., González-Galindo, F., Yiğit, E., Feofilov, A.G., Forget, F., Hartogh, P., 2015. Cooling of the Martian thermosphere by $CO_2$ radiation and gravity waves: An intercomparison study with two general circulation models. J. Geophys. Res. Planets 120, 913–927.

Medvedev, A.S., Yiğit, E., 2019. Gravity Waves in Planetary Atmospheres: Their Effects and Parameterization in Global Circulation Models. Atmosphere 10, 531.

Medvedev, A.S., Yiğit, E., Hartogh, P., 2011a. Estimates of gravity wave drag on Mars: Indication of a possible lower thermospheric wind reversal. Icarus 211, 909–912.

Medvedev, A.S., Yiğit, E., Hartogh, P., Becker, E., 2011b. Influence of gravity waves on the Martian atmosphere: General circulation modeling. J. Geophys. Res. 116, E10004.

Milby, Z., de Kleer, K., Schmidt, C., Leblanc, F., 2024. Short-timescale spatial variability of Ganymede's optical Aurora. Planet. Sci. J. 5, 153.

Moore, C.H., Goldstein, D.B., Varghese, P.L., Trafton, L.M., Stewart, B., 2009. 1-D DSMC simulation of Io's atmospheric collapse and reformation during and after eclipse. Icarus 201, 585–597.

Parish, H.F., Schubert, G., Hickey, M.P., Walterscheid, R.L., 2009. Propagation of tropospheric gravity waves into the upper atmosphere of Mars. Icarus 203, 28–37.

Rhode, S., Preusse, P., Ungermann, J., Polichtchouk, I., Sato, K., Watanabe, S., Ern, M., Nogai, K., Sinnhuber, B.-M., Riese, M., 2024. Global-scale gravity wave analysis methodology for the ESA Earth Explorer 11 candidate CAIRT. Atmos. Meas. Tech. 17, 5785–5819.

Roeten, K.J., Bougher, S.W., Yiğit, E., Medvedev, A.S., Benna, M., Elrod, M.K., 2022. Impacts of gravity waves in the martian thermosphere: The mars global ionosphere-thermosphere model coupled with a whole atmosphere gravity wave scheme. J. Geophys. Res. Planets 127. https://doi.org/10.1029/2022je007477

Schubert, G., Hickey, M.P., Walterscheid, R.L., 2005. Physical processes in acoustic wave heating of the thermosphere. J. Geophys. Res. 110. https://doi.org/10.1029/2004jd005488

Shematovich, V., Johnson, R., Cooper, J., Wong, M., 2005. Surface-bounded atmosphere of Europa. Icarus 173, 480–498.

Siddle, A.G., Mueller-Wodarg, I.C.F., Stone, S.W., Yelle, R.V., 2019. Global characteristics of gravity waves in the upper atmosphere of Mars as measured by MAVEN/NGIMS. Icarus 333, 12–21.

Terada, N., Leblanc, F., Nakagawa, H., Medvedev, A.S., Yiğit, E., Kuroda, T., Hara, T., England, S.L., Fujiwara, H., Terada, K., Seki, K., Mahaffy, P.R., Elrod, M., Benna, M., Grebowsky, J., Jakosky, B.M., 2017. Global distribution and parameter dependences of gravity wave activity in the Martian upper thermosphere derived from MAVEN/NGIMS observations. J. Geophys. Res. Space Phys. 122, 2374–2397.

Tian, L., Johnson, R.E., Tucker, O.J., Woodson, A.K., Williamson, H.N., Mogan, S.R.C., 2023. Molecular Kinetic Simulations of Transient and Steady Wave Propagation into a Planet's Exosphere. Atmosphere 14, 441.

Tolson, R.H., Keating, G.M., Zurek, R.W., Bougher, S.W., Justus, C.G., Fritts, D.C., 2007. Application of Accelerometer Data to Atmospheric Modeling During Mars Aerobraking Operations. J. Spacecr. Rockets 44, 1172–1179.

Tseng, W.-L., Lai, I.-L., Hsu, H.-W., Ip, W.-H., Wu, J.-S., 2025. Surface deposition of icy dust entrained in Europa's plumes. Planet. Sci. J. 6, 90.

Tseng, W.-L., Lai, I.-L., Ip, W.-H., Hsu, H.-W., Wu, J.-S., 2022. The 3D Direct Simulation Monte Carlo study of Europa's gas plume. Universe 8, 261.

Tucker, O.J., Combi, M.R., Tenishev, V.M., 2015. 2D models of gas flow and ice grain acceleration in Enceladus' vents using DSMC methods. Icarus 257, 362–376.

Tucker, O.J., Erwin, J.T., Deighan, J.I., Volkov, A.N., Johnson, R.E., 2012. Thermally driven escape from Pluto's atmosphere: A combined fluid/kinetic model. Icarus 217, 408–415.

Tucker, O.J., Johnson, R.E., Deighan, J.I., Volkov, A.N., 2013. Diffusion and thermal escape of H2 from Titan's atmosphere: Monte Carlo simulations. Icarus 222, 149–158.

Tucker, O.J., Killen, R.M., Johnson, R.E., Saxena, P., 2021. Lifetime of a transient atmosphere produced by lunar volcanism. Icarus 359, 114304.

Tucker, O.J., Waalkes, W., Tenishev, V.M., Johnson, R.E., Bieler, A., Combi, M.R., Nagy, A.F., 2016. Examining the exobase approximation: DSMC models of Titan's upper atmosphere. Icarus 272, 290–300.

Volkov, A.N., Tucker, O.J., Erwin, J.T., Johnson, R.E., 2011. Kinetic simulations of thermal escape from a single component atmosphere. Phys. Fluids 23, 066601.

Waite, J.H., Jr, Greathouse, T.K., Carberry Mogan, S.R., Sulaiman, A.H., Valek, P., Allegrini, F., Ebert, R.W., Gladstone, G.R., Kurth, W.S., Connerney, J.E.P., Clark, G., Bagenal, F., Duling, S., Romanelli, N., Bolton, S., Vorburger, A., Paranicas, C., Kollmann, P., Mauk, B., Hansen, C., Buccino, D., Johnson, R.E., Wilson, R.J., Teolis, B., 2024. Magnetospheric-ionospheric-atmospheric implications from the Juno flyby of Ganymede. J. Geophys. Res. Planets 129. https://doi.org/10.1029/2023je007859

Walterscheid, R.L., Hickey, M.P., Schubert, G., 2013. Wave heating and Jeans escape in the Martian upper atmosphere. J. Geophys. Res. Planets 118, 2413–2422.

Wang, X., Lian, Y., Cui, J., Richardson, M., Wu, Z., Li, J., 2020. Temperature variability in titan's upper atmosphere: The role of wave dissipation. J. Geophys. Res. Planets 125. https://doi.org/10.1029/2019je006163

Williamson, H.N., Johnson, R.E., Leclercq, L., Elrod, M.K., 2019. Large amplitude perturbations in the martian exosphere seen in MAVEN NGIMS data. Icarus 331, 110–115.

Withers, P., 2006. Mars Global Surveyor and Mars Odyssey Accelerometer observations of the Martian upper atmosphere during aerobraking. Geophys. Res. Lett. 33, L02201.

Wright, C.J., 2012. A one-year seasonal analysis of martian gravity waves using MCS data. Icarus 219, 274–282.

Yiğit, E., 2023. Coupling and interactions across the Martian whole atmosphere system. Nat. Geosci. 16, 123–132.

Yiğit, E., 2021. Martian water escape and internal waves. Science 374, 1323–1324.

Yiğit, E., Aylward, A.D., Medvedev, A.S., 2008. Parameterization of the effects of vertically propagating gravity waves for thermosphere general circulation models: Sensitivity study. J. Geophys. Res. 113. https://doi.org/10.1029/2008jd010135

Yiğit, E., England, S.L., Liu, G., Medvedev, A.S., Mahaffy, P.R., Kuroda, T., Jakosky, B.M., 2015. High-altitude gravity waves in the Martian thermosphere observed by MAVEN/NGIMS and modeled by a gravity wave scheme. Geophys. Res. Lett. 42, 8993–9000.

Yiğit, E., Medvedev, A.S., 2015. Internal wave coupling processes in Earth's atmosphere. Adv. Space Res. 55, 983–1003.

Yiğit, E., Medvedev, A.S., Benna, M., Jakosky, B.M., 2021. Dust storm-enhanced gravity wave activity in the martian thermosphere observed by MAVEN and implication for atmospheric escape. Geophys. Res. Lett. 48. https://doi.org/10.1029/2020gl092095

## Appendix A1 Extended Cycle Analysis for Wave Pattern Convergence

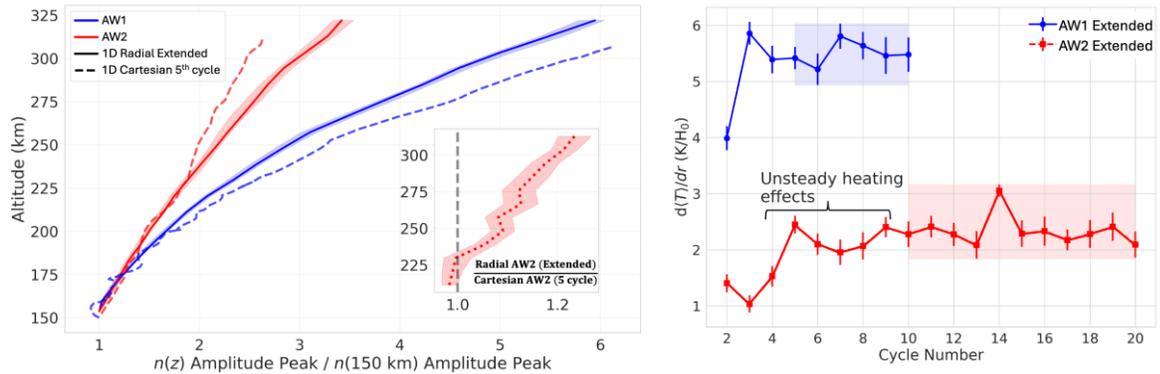

*Fig. A1. Left* panel: *Density amplitude peak growth versus altitude for extended radial (solid) and the five cycle Cartesian (dashed) simulations of AW1 (blue) and AW2 (red), normalized by their respective amplitudes at 150 km. Solid lines represent cycle-averaged amplitude peak profiles: AW1 averaged over cycles 5-10, AW2 averaged over cycles 10-20. Shaded bands indicate 2σ confidence intervals reflecting cycle-to-cycle variability at each altitude. Inset: AW2 radial/Cartesian ratio of amplitude peak growth. Right: Cycle-average temperature gradients $d\langle T \rangle/dr$ for both wave modes, computed as in Sec. 3.3, across all cycles. Error bars represent standard error of the linear fit slope. Colored rectangles mark the cycle ranges used for averaging the peak growth curves.*

Two additional sets of simulation runs were conducted to explore unsteady perturbations observed during the initial five-cycle analysis assumed in T23 to verify convergence of the wave-averaged temperature gradient $d\langle T \rangle/dr$. In DSMC, steady state can require significant run times to obtain a steady state result. As shown in Fig. A1, for AW2 the simulated heating gradient is unsteady up to 9 cycles after onset. To characterize convergence behavior on the amplitude growth and heating effects, we carried out additional simulations including 10 cycles for AW1 and 20 cycles for AW2.

The lower-frequency AW1 mode stabilizes to $d\langle T \rangle/dr \sim 5.6$ K/$H_0$ by the fifth cycle, confirming the 40% reduction relative to T23's Cartesian value of 9.4 K/$H_0$. The higher-frequency AW2 mode reaches five cycles in half that time and exhibits convergence only after cycle 10. The converged value is $d\langle T \rangle/dr = 2.34 \pm 0.26$ K/$H_0$ compared to our value of 1.94 K/$H_0$ obtained at the 5th cycle. This value represents a 47% reduction from T23's Cartesian result of 4.4 K/$H_0$. When excluding cycle 14 which is an outlier, the value becomes $2.26 \pm 0.11$ K/$H_0$ corresponding to a 49% reduction to the Cartesian analog. The five-cycle analysis presented in Sect. 3.3 identified a heating reduction of 56% relative to the Cartesian analog. Similarly, the extended AW2 simulation reveals that density amplitude growth changes upon convergence, reaching values ~20% higher than T23 at 320 km altitude, as seen in the left panel of Fig. 1A. Direct comparison between the fifth-cycle results in Fig. 3 must be interpreted cautiously for AW2, as neither the 1D radial nor Cartesian simulations had converged at this stage. The extended simulations (Fig. 1A) demonstrate that both the amplitude peak growth and cycle-averaged temperature gradient change after five cycles for AW2, confirming conclusions about geometric effects require converged wave solutions. The ~20% increased amplitude peak growth in the extended AW2 run reflects the ongoing thermalization process rather than a fundamental geometric difference from the Cartesian case. While the extended steady forcing over 10-20 cycles is not physically realistic in Mars' dynamic atmosphere, these conditions were solely used to compare to steady state solutions. More realistic time dependent studies will be considered in future work.